\documentclass[manuscript]{aastex}
\usepackage{emulateapj5}
\usepackage{apjfonts}
\usepackage{epsfig}
\usepackage{lscape}

\journalinfo{The Astrophysical Journal Letters in press}
\slugcomment{Received 2007 December 6; accepted 2008 February 21}
\shorttitle{Constraints on Spins of Supermassive Black Hole in M87}
\shortauthors{WANG ET AL.}

\def\rd{{\rm d}}

\def\ergs{\ifmmode {\rm erg~ s^{-1}} \else {\rm erg~s^{-1}}\ \fi}
\def\kms{\ifmmode {\rm km~ s^{-1}} \else {\rm km~s^{-1}}\ \fi}

\def\mbh{M_{\bullet}}

\def\mgii{\ifmmode Mg {\sc ii} \else Mg {\sc ii}\ \fi}

\def\sunm{M_{\odot}}
\def\taugg{\tau_{\gamma\gamma}}

\def\lax{{$\mathrel{\hbox{\rlap{\hbox{\lower4pt\hbox{$\sim$}}}\hbox{$<$}}}$}}
\def\gax{{$\mathrel{\hbox{\rlap{\hbox{\lower4pt\hbox{$\sim$}}}\hbox{$>$}}}$}}

\def\ergs{${\rm erg~s^{-1}}$}

\begin{document}

\title{Spins of the supermassive black hole in M87: new constraints from TeV observations}

\author{Jian-Min Wang\altaffilmark{1,2},
        Yan-Rong Li\altaffilmark{1},
	Jian-Cheng Wang\altaffilmark{3}
	and Shu Zhang\altaffilmark{1}}

\altaffiltext{1}
{Key Laboratory for Particle Astrophysics, Institute of High Energy Physics,
CAS, 19B Yuquan Road, Beijing 100049, China}

\altaffiltext{2}
{Theoretical Physics Center for Science Facilities (TPCSF), CAS}

\altaffiltext{3}
{Yunnan Observatory, CAS, Kunming 650011, China}

\begin{abstract}
The rapid TeV $\gamma-$ray variability detected in the well-known nearby radio galaxy M87 implies 
an extremely compact emission region ($5-10$ Schwarzschild radii) near the horizon of the supermassive 
black hole in the galactic center. TeV photons are affected by dilution due to interaction with the 
radiation field of the advection-dominated accretion flow (ADAF) around the black hole, and can thus 
be used to probe the innermost regions around the black hole.  We calculate the optical depth of the 
ADAF radiation field to the TeV photons and find it strongly depends on the spin of the black hole. 
We find that transparent radii of 10 TeV photons are of $5R_{\rm S}$ and $13R_{\rm S}$ for the maximally
rotating and non-rotating black holes, respectively. With the observations, the calculated transparent 
radii strongly suggest the black hole is spinning fast in the galaxy. TeV photons could be used as a 
powerful diagnostic for estimating black hole spins in galaxies in the future.
\end{abstract}

\section{Introduction}
It is generally accepted that accretion onto supermassive black holes (SMBHs) is the major 
source of their growth during the periods of the active nuclei duty cycles (Chokshi \& Turner 1992; 
Yu \& Tremaine 2002; Marconi et al. 2004; Wang et al. 2008). This process spins up the SMBHs, 
resulting in that the most SMBHs should be rapidly rotating (Bardeen 1972; Thorne 1974; Volonteri 
et al. 2005; Wang et al. 2006). The only strong evidence for rapidly rotating individual extragalactic 
SMBHs is the gravitationally broaden profile of the iron K$\alpha$ emission line in the X-ray spectrum 
of MCG-6-30-15 (Wilms et al. 2001; Fabian et al. 2002; Brenneman \& Reynolds 2006). The general lack 
of observational evidence for gravitationally broaden iron K$\alpha$ lines leaves the issue of black 
hole spin as one of the most elusive questions in astrophysics.

It is well-known that M87 (at a distance of $\sim 16$ Mpc) contains an SMBH with a mass of 
$\mbh=(3.2\pm 0.9)\times 10^9\sunm$ based on observations with {\em Hubble Space Telescope} (Harm et 
al. 1994; Macchetto et al. 1997). Thanks to the high spatial resolution of these observations, structures 
in the core of M87 have been resolved on scales of 20pc at optical wavelengths (Harm et al. 1994) and 
$\sim 100R_{\rm S}$ in radio wavelengths (Junor et al. 1999), where $R_{\rm S}=2G\mbh/c^2$ is the 
Schwarzschild radius, $G$ the gravitational constant and $c$ the light speed. The multiwavelength
continuum can be explained by a simple ADAF or modified model with outflows under a Schwarzschild hole 
(Di Matteo et al. 2000; 2003). Several parameters in these models are strongly degenerated,
it remains an open question whether the SMBH is rapidly rotating. 

The goal of the present paper is to estimate the SMBH spin based on the TeV variability detected in M87 
by High Energy Stereoscopic System (HESS) (Aharonian et al. 2006). An origin from jet has been ruled out 
(Aharonian et al. 2006). The attenuation of TeV photons by the ADAF radiation field depends on the 
spectral energy distribution (SED) along the ADAF radius. This presents us with the opportunity to tackle 
this issue. 

\section{Advection-dominated accretion flow in M87}
Fig. 1 shows the multiwavelength continuum of M87. From radio to X-rays, the SED is dominated by emission 
from the accretion flow around the SMBH (Di Matteo et al. 2003). The spectrum has a bolometric 
luminosity of $\sim 10^{41}$\ergs, giving the accretion flow an Eddington ratio of $\sim 10^{-6}$. This
implies that an advection-dominated accretion flow is at work in the galaxy (Reynolds et al. 1996; Di 
Matteo et al. 2003). For simplicity, we use a self-similar solution of the ADAF (Narayan \& Yi 1994) to 
determine the temperatures of both protons and electrons (Narayan \& Yi 1995) in order to calculate the 
emergent spectrum. The self-similar solution of the ADAF is characterized by a factor $f$, which describes 
the fraction of the released gravitational energy that is advected into the black hole. The tendency of 
$f\rightarrow 1$ has been verified by extensive numerical calculations of the global solution of the ADAF 
(e.g. Narayan et al. 1997; Manmoto et al. 1997). We define $\dot{m}=\eta\dot{M}c^2/L_{\rm Edd}$, where 
$\dot{M}$ is the accretion rate, $\eta(=0.1)$ the radiative efficiency and $L_{\rm Edd}$ the Eddington 
luminosity. There are five parameters ($\alpha,\beta,\dot{m},\mbh, R_{\rm in}$) in the ADAF model, where 
$\alpha$ is the viscosity, $R_{\rm in}$ the inner radius of the ADAF, 
$\beta=P_{\rm gas}/\left(P_{\rm gas}+P_{\rm B}\right)$, $P_{\rm gas}$ and $P_{\rm B}$ are the gas and 
magnetic pressures, respectively. We refer Schwarzschild and Kerr holes to those with the 
specific angular momentum $a=0$ and $a=1$, respectively, where $a=J/G\mbh^2 c^{-1}$ and $J$ is the
angular momentum of the hole. Spins are directly related to the inner radius $R_{\rm in}$ , which can be 
determined by fitting observed continuum. For Schwarzschild and Kerr holes, the inner radius 
$R_{\rm in}=6R_{\rm g}$ and $R_{\rm g}$, respectively, where $R_{\rm g}=R_{\rm S}/2$ is the gravitational 
radius. We begin by fitting the observed multiwavelength SED of M87 to determine the best-fit parameters 
of the accretion flow. Then, we find the optical depth of the ADAF radiation to TeV photons. We include 
synchrotron, multiple Compton scattering and free-free emission processes in this calculations. We follow 
the numerical scheme for the emergent spectrum described in Manmoto et al. (1997) (see also Yuan et al. 
2000) in this Letter.

\begin{figure*}[t]
\centerline{\includegraphics[angle=-90,width=15.50cm]{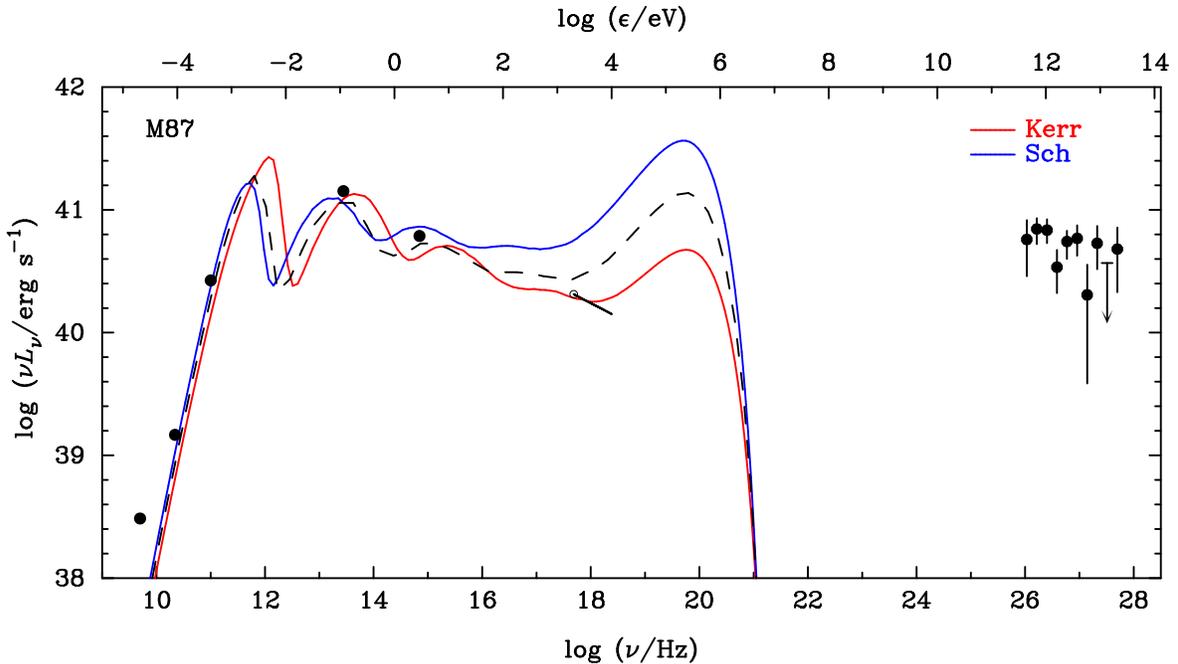}}
\figcaption{\footnotesize
The spectral energy distribution of M87 from radio to TeV bands.  The three radio data points are from VLBI 
nuclear flux density measurements (Pauliny-Toth et al. 1981; Spencer et al. 1986; B\"a\"ath et al. 1992 in 
increasing frequency order). The near IR 10$\mu$ nuclear flux measurement is the best IR flux limit 
(a resolution of $\sim 0^{\prime\prime}.46$) from {\em Gemini} observations (Perlman et al. 2001). The 
optical nuclear continuum measurement is from {\em HST} observations with a resolution of $0^{\prime\prime}.15$ 
(Harms et al. 1994). The black line with an open circle is the {\em Chandra} spectrum ($0.2-10$ keV) (Di Matteo 
et al. 2003). The color lines are different ADAF models for Kerr (red) and Schwarzschild (blue) black holes. 
The black dashed line is the spectrum from ADAF model with $a_{\rm TeV}=0.65$ and $\dot{m}=1.1\times 10^{-3}$.
}
\label{fig1}
\end{figure*}

We used the black hole mass measured by {\em HST} in fitting the continuum. The viscosity and advection factors 
of $\alpha$ and $f$ are actually absorbed into one parameter in the self-similar solution. Though the viscosity 
$\alpha$ is highly unknown, only their combination of $\alpha$ and $f$ is important for the SED model (Narayan 
\& Yi 1995). The final optical depth to TeV photons is directly dependent on the radial SED rather than 
$\alpha$ and $f$, so we take typical values of $\alpha=0.3$ and $f=0.99$.  The ADAF spectrum is 
characterized by a sharp peak of synchrotron emission from moderately relativistic electrons, multiple bumps 
from inverse Compton scattering and a high energy peak with a cutoff originating from bremsstrahlung emission
from the very hot plasma in the ADAF. The roles of $\beta$ and $\dot{m}$ are clear in fittings of the
continuum (Narayan \& Yi 1995) if the inner radius of the ADAF is fixed. For a plasma with equipartition
between the gas and magnetic fields, $\beta_{\rm eq}=0.5$. For $\beta>\beta_{\rm eq}$, the plasma
is weakly magnetized and for $\beta<\beta_{\rm eq}$ the plasma is strongly magnetized. Usually $\beta=0.5$ works
generally, but it is needed to be elaborately adjusted for specific fittings. For a model with a fixed 
$R_{\rm in}$, the synchrotron peak is primarily determined by $\mbh$, while the Compton bumps and X-ray peak are
sensitive to $\dot{m}$ and $\beta$. Fig. 1 shows the best fittings ADAF models with $R_{\rm in}=1,6R_{\rm g}$. 
The Kerr model,  with $\beta=0.35$ and $\dot{m}=5.5\times 10^{-4}$, is consistent with the entire spectrum from 
radio to X-rays, while the Schwarzschild model with $\beta=0.35$ and $\dot{m}=2.0\times 10^{-3}$ can match the 
data at only radio, near IR and optical wavelengths, but always not with the X-ray data. The differences between 
the two models originate from the fact that the Kerr black hole has higher radiation efficiency than the 
Schwarzschild. For the same total luminosity, the accretion rate is lower for the Kerr hole than
the Schwarzschild, resulting in a thinner medium of the accreting gas and then reduced free-free emission in
X-ray band. We note, however, that modified models of the ADAF might in principle produce a variety of 
spectra (Di Matteo et al. 2003) and that measurements of the SMBH spins strongly depend upon $\dot{m}$. 
We present a new method for revealing the SMBH spin using the constraints from TeV variability.

Uncertainties in the ADAF fit model results are primarily governed by $\dot{m}$ and $R_{\rm in}$ for models 
with fixed $\alpha$ and $f$. In particular, the modeled X-ray spectrum strongly depends on both $\dot{m}$ and 
$R_{\rm in}$. It is difficult to estimate these uncertainties, however the upper limit of $R_{\rm in}$ will be 
constrained by the upper limit of the region responsible for the observed TeV variability.

%\vglue 0.7cm
\figurenum{2}
\centerline{\psfig{figure=f2.ps,angle=270,width=7.5cm}}
\figcaption{\footnotesize 
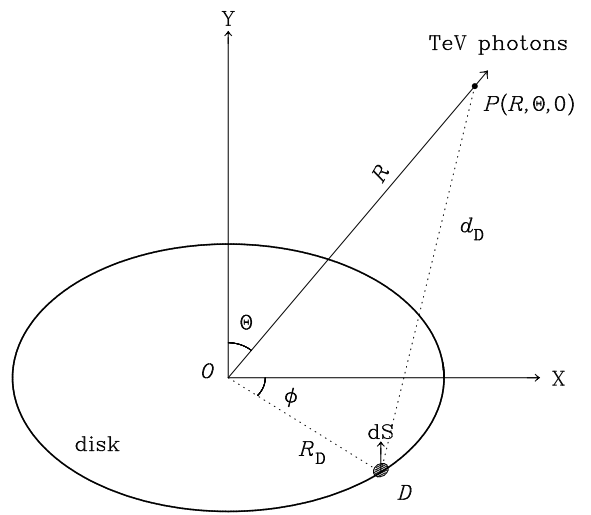
The geometric scheme. For simplicity, we assume that the
TeV photons originate from a point $P$ in the XOY$-$plane. $R$ is the distance of the TeV 
photons to the hole, $R_{\rm D}$ is the radius of the ADAF, $\Theta$ is the viewing angle.
}
\label{fig2}
\vglue 0.5cm

\section{Optical depth of TeV photons and hole's spins}
We calculate the number density of soft photons from the ADAF as a function of position $P(R,\Theta,0)$. Fig. 2 
shows the geometric scheme for the ADAF disk. The vertical structure of the ADAF is unknown, but it is generally 
assumed that the vertical density has an exponential profile (e.g. Manmoto et al. 1997). We follow this with the 
assumption that most of the emitted photons originate from the mid-plane of the accretion disk. With simple 
manipulation, we have $\cos({\rm POD})=\sin\Theta\cos\phi$, and the distance
\begin{equation}
d_{\rm D}^2=R^2+R_{\rm D}^2-2RR_{\rm D}\sin\Theta\cos\phi.
\end{equation}
The cosine of the angle between two interacting photons is then
\begin{equation}
\mu=\frac{R-R_{\rm D}\sin \Theta\cos \phi}{d_{\rm D}}.
\end{equation}
We then have the number density of photons from the ADAF, given by
\begin{equation}
n_{\rm ph}(\Theta,R,\epsilon,R_{\rm D})=\frac{F(R_{\rm D},\epsilon)}{2\pi d_{\rm D}^2\epsilon m_ec^3}
\left(\frac{R}{d_{\rm D}}\right)\cos\Theta ,
\end{equation}
where $F(R_{\rm D},\epsilon)$ is the upper stream flux of photons from a radius $R_{\rm D}$,
the factor of $\left(R/d_{\rm D}\right)\cos\Theta$ is the projected area element in the direction
DP, $m_e$ is the mass of the electron and $\epsilon$ is the photon energy in units of $m_ec^2$.

\figurenum{3}
\centerline{\psfig{figure=f3.ps,angle=270,width=7.5cm}}
\figcaption{\footnotesize 
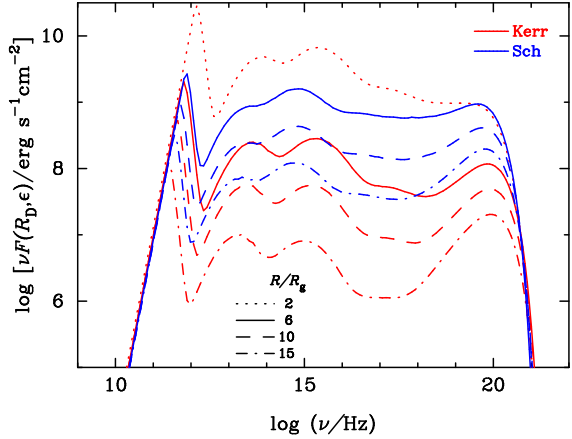
The SEDs at 4 and 3 different radii for Kerr (red) and Schwarzschild (blue) black holes with the
parameters obtained from the continuum fitting, respectively. 
}
\label{fig3}
\vglue 0.3cm

Fig. 3 shows the SEDs for both Kerr and Schwarzschild black holes at four and three different radii, 
respectively. The SEDs demonstrate that gravitational energy is dissipated in a more compact region for 
a Kerr hole than a Schwarzschild. This is the key point that provides an opportunity to probe the black 
hole spin. We stress here that we neglect general relativistic (GR) effects in this paper, however the 
release of gravitational energy via viscous dissipation will be concentrated in a more compact region 
if we include the effects. This, in fact, increases the robustness of the conclusions presented here.
More sophisticated model including GR effects will be given in a future paper.

TeV photons can divulge details of the ambient radiation field through their dilution via 
pair production. Pair production can not be avoided if energies of two colliding photons 
$\epsilon_1$ and $\epsilon_2$ (in units of $m_ec^2$) satisfy the relation 
\begin{equation}
\epsilon_1\epsilon_2=\frac{2}{(1-\mu)(1-v^2)},
\end{equation}
where $\mu$ is the cosine of the angle between the two photons, and $v$ is the velocity of the positrons 
and electrons in the center of momentum frame in the units of $c$. The optical depth of the radiation 
field to TeV photons can be obtained by integrating over the radius $R$ along 
the propagation path of the TeV photons to infinity and direction of incident soft photons, i.e.,
\begin{equation}
\begin{array}{r}
\taugg(\Theta,R,\epsilon)=\displaystyle\int_R^{\infty}\rd R\int_{R_{\rm in}}^{R_{\rm out}}R_{\rm D}\rd R_{\rm D} 
                          \int_0^{2\pi}\rd\phi\int_{\epsilon_t} \rd\epsilon^{\prime}\\
      \\
    \displaystyle\sigma_{\gamma\gamma}
(\epsilon,\epsilon^{\prime},\mu) n_{\rm ph}(\Theta, R,\epsilon^{\prime},R_{\rm D}),
\end{array} 
\end{equation}
where $\sigma_{\gamma\gamma}(\epsilon,\epsilon^{\prime},\mu)$ is the cross section of the two
colliding photons (Gould \& Schr\'eder 1967), $\epsilon_t$ is the threshold energy of the interacting photons
and $R_{\rm out}$ is the outer radius of the ADAF. It should be noted that $\taugg$ strongly depends on the
incident angle between the two photons. 

The dependence of $\taugg$ on $R_{\rm in}$ (and thus the SMBH spin) can be qualitatively predicted. 
For a given bolometric luminosity, most of the gravitational energy dissipated during accretion onto a 
rapidly black hole is released in a more compact region than in a system containing a Schwarzschild black 
hole. This leads to a much steeper $\taugg-R$ relation around the Kerr black hole horizon and facilitates 
the escape of TeV photons from considerable smaller radii. It is thus expected that the timing information 
from TeV photons can be used as a diagnostic of the spatial distributions of the soft photon field surrounding 
the SMBH and thus its spin. 

\figurenum{4}
\centerline{\psfig{figure=f4.ps,angle=270,width=7.50cm}}
\figcaption{\footnotesize 
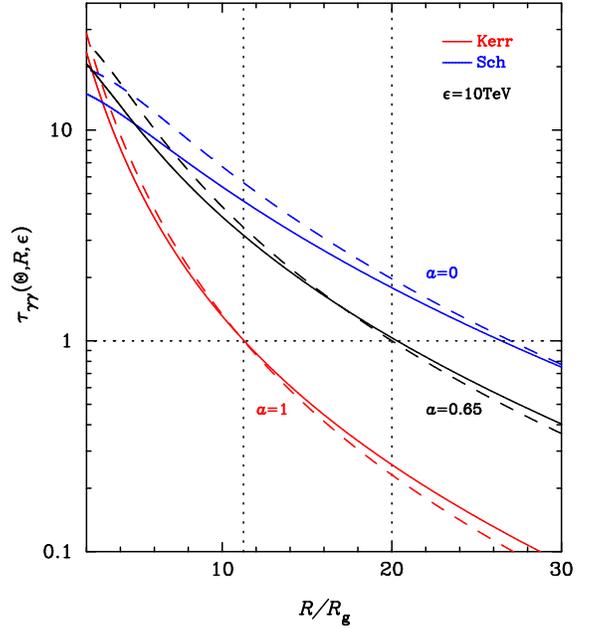
The optical depth of the ADAF radiation field to 10 TeV photons. The solid and dashed lines are for 
$\Theta=0^{\circ}$ and $30^{\circ}$, respectively. The radio jet has an orientation angle of $30^{\circ}$
with respect to an observer's sight line, so we assume the ADAF is viewed at $\Theta=30^{\circ}$. We only 
plot $\taugg$ to $10$TeV photons, because $\taugg\le 1$ for most radii for photons $<1$TeV. 
}
\label{fig4}
\vglue 0.5cm

Fig. 4 shows the optical depth of the radiation field to 10 TeV photons for the best fitting ADAF 
models for both Schwarzschild and Kerr black holes. We fix $R_{\rm out}=10^3R_{\rm g}$. We note that
the final results are not sensitive to $R_{\rm out}$. $\taugg$ is sensitive to radii, but not
to the viewing angle. As we argue qualitatively, Fig. 4 shows that the $\taugg-R$ relation is much steeper
for Kerr holes than the Schwarzschild. For TeV photons to be able to escape from the radiation field 
of the ADAF, we require $\taugg\le 1$. We define the transparent radius,  $R_c$, as the radius at which
$\taugg=1$. We find $R_c\approx 11R_{\rm g}$ and $26R_{\rm g}$ for Kerr and Schwarzschild black holes, 
respectively. The TeV variability timescale in M87, observed by HESS, is of $\sim 2$days, implying 
a compact emission region of $R_{\rm TeV}\le c{\cal D}\Delta t_{\rm TeV}\approx 10~{\cal D}R_{\rm g}$, 
where ${\cal D}=1/\Gamma\left(1-V\cos\Theta/c\right)$ is the Doppler factor and 
$\Gamma=\left[1-\left(V/c\right)^2\right]^{-1/2}$ the Lorentz factor of a jet with a velocity of $V$. 
If the jet is viewed at $\Theta=30^{\circ}$ (Bicknell \& Begelman 1996; Aharonian et al. 2006), the maximum 
Doppler factor is ${\cal D}_{\rm max}=2$. Even if the TeV photons are emitted from the base of the relativistic 
jet, we have $R_{\rm TeV}\le 20~{\cal D}_2R_{\rm g}$, where ${\cal D}_2={\cal D}_{\rm max}/2$. On the necessary
condition of $R_{\rm TeV}\ge R_c$, the HESS observations suggest the presence of a rapidly rotating black hole 
in M87 in term of $R_{\rm c}=26R_{\rm g}>R_{\rm TeV}$ for a Schwarzschild hole. Otherwise, 
the observed TeV region would be optical thick to 10TeV. Increasing the black hole spin, $a$, the transparent
radius $R_c$ decreases. In order for the transparent radius to equal the TeV emission radius, we require a black 
hole spin of $a_{\rm TeV}=0.65$ for $\dot{m}=1.1\times 10^{-3}$. We note that the X-ray 
spectrum can not be well fit with $a=0.65$ and $\dot{m}=1.1\times 10^{-3}$, however we show the SED in Figs 1 
and 4 to demonstrate how sensitive the transparent radius and the SED are to black hole spin. Fig 4 shows a 
region between $1\ge a\ge a_{\rm TeV}$, in which the exact spins should be, most likely have the maximum spin
if the TeV region has a radius of $10R_{\rm g}$. 

We note that although the $\taugg-R$ relation becomes steeper with spin, $R_{\rm c}$ is in fact insensitive to 
spins of $a<a_{\rm TeV}$, making this technique less useful for estimation of their spins in slowly rotating
black holes. The absence of any evidence for a potential break between $10-20$TeV means $\taugg\ll 1$ for 
these TeV photons, namely the spin $a>a_{\rm TeV}$. Otherwise, at $R=10R_{\rm g}$ for a Schwarzschild hole, 
$\taugg=5\sim 6$ and the TeV photons will be attenuated by a factor of $e^{\taugg}\approx 150\sim 400$. 
This strongly conflicts with results from the TeV observations. The conclusion does, however, depend on the 
fitting of the multiwavelength continuum, which could be contaminated by other sources, for example, the inner 
jet. The current data and model do not allow us to constrain the spin to better than $a>0.65$, however, future 
improved multiwavelength data and models will produce tighter constraints on how fast the spin is. 

A rapid rotation is generally consistent with the statistic results from the Sloan Digital Sky Survey (Wang 
et al. 2006) and with justification from radio emission in term of Blandford-Znajek effect (Newman et al. 2007). 
Since SMBH growth is primarily accretion, the spin angular momentum of the black hole is likely to be
dominated by its accretion history. If the accretion is random, however, as suggested by King \& Pringle (2007), 
the net spins of SMBHs should tend to zero. The spin evolution of SMBHs will thus uncover the accretion modes 
(co-rotating, retrograde or random accretion) in their history from measuring spins of local SMBHs.

\section{Conclusions and discussions}
The TeV variability as a new independent ingredient allows us to disentangle the inner radius from other parameters 
in modeling the multiwavelength continuum. We fit the multiwavelength continuum of M87 and use the best-fit parameters 
of the ADAF to calculate the optical depth of the ADAF radiation field to TeV photons.  This optical depth is sensitive 
to the spin of SMBHs and combined with the TeV variability constraints, we conclude that the SMBH must be rapidly 
rotating in M87. Whatever mechanism produces the TeV photons in M87, these energetic photons provide a potentially 
powerful probe of the SMBH spin. 

We should note some uncertainties of the present results, which fully depends on the theoretical model of ADAF, and
contamination of the radiation in the central region from its host and jet. The influence of potential outflows from 
the ADAF should be examined in a future paper. Although we neglect the GR effects and the global structure of the 
ADAF in this paper for simplicity, the addition of the effects will make the conclusion of this work robust, because 
the gravitational energy is dissipated in more compact regions due to the deeper gravitational potential than in the 
Newtonian case. The resulting $\taugg-R$ relation will then be steeper, reducing the transparent radius from which 
TeV photons can escape. Accurate measurements of the black hole spin rely on global calculations of the ADAF with 
general relativity and the timescale measurements of TeV variability to confine the dimensions of the TeV region. 
Future observations of other sources with HESS may reveal similar TeV variability to M87, particularly in low 
luminosity active galactic nuclei (LLAGNs), such as radio galaxies and low ionization nuclear emission regions 
(LINERs). This would lead to further estimates of the spins of SMBHs in nearby galaxies. Additionally, the present 
results suggest that any mechanism responsible for production of the TeV photons should be located outside a region 
with $\taugg\ge 1$, constraining on the radiation mechanism of TeV photons.

\acknowledgements{We are grateful to the referee for a helpful report.
N. Schurch is acknowledged for careful reading the manuscript and helpful comments.
J.M.W. thanks F. Yuan for his kind help in calculations of emergent spectrum from the ADAF.
We appreciate the stimulating discussions among the members of IHEP AGN 
group. The research is supported by NSFC and CAS via NSFC-10325313, 10733010 and 
10521001, and KJCX2-YW-T03, respectively.}

\end{document}